\documentclass[default,grl]{AGUTeX}

\usepackage[maxfloats=25]{morefloats}

\usepackage{graphicx}

\setkeys{Gin}{draft=false}

\authorrunninghead{DESAI ET AL.}

\titlerunninghead{Pickup ions at Rhea}

\authoraddr{Corresponding author: R. T. Desai, Blackett Laboratory, Imperial College London, London, UK. (ravindra.desai@imperial.ac.uk)}

\usepackage{amsmath}
\usepackage{amssymb}

\begin{document}


\title{Cassini CAPS identification of pickup ion compositions at Rhea}




\authors{
R. T. Desai\altaffilmark{1,2,3}, 
S. A. Taylor\altaffilmark{1,2},
L. H. Regoli\altaffilmark{4}, 
A. J. Coates\altaffilmark{1,2},
T. A. Nordheim\altaffilmark{5}, 
M. A. Cordiner\altaffilmark{6}, 
B. D. Teolis\altaffilmark{7}, 
M. F. Thomsen\altaffilmark{8}, 
R. E. Johnson\altaffilmark{9}, 
G. H. Jones\altaffilmark{1,2}, 
M. M. Cowee\altaffilmark{10}, 
J. H. Waite\altaffilmark{7}
}
\altaffiltext{1}{Mullard Space Science Laboratory, 
University College London, UK.}
 \altaffiltext{2}{Centre for Planetary Science at UCL/Birkbeck, London, UK.}
 \altaffiltext{3}{Blackett Laboratory, Imperial College London, UK.}
\altaffiltext{4}{Michigan University, Department of Space Science and Engineering, Michigan, USA.}
\altaffiltext{5}{NASA-JPL/Caltech, Pasadena, California, USA.}
\altaffiltext{6}{NASA Goddard Space Flight Center, Maryland, USA.}
\altaffiltext{7}{Space Science and Engineering Division, Southwest Research Institute, Texas, USA.}
\altaffiltext{8}{Planetary Science Institute, Tucson, Arizona, USA.}
\altaffiltext{9}{Engineering Physics, University of Virginia, Charlottesville, USA.}
\altaffiltext{10}{Los Alamos National Laboratory, Los Alamos, New Mexico, USA.}





\begin{abstract}

Saturn's largest icy moon, Rhea, hosts a tenuous surface-sputtered exosphere composed primarily of molecular oxygen and carbon dioxide. In this Letter, we examine Cassini Plasma Spectrometer velocity space distributions near Rhea and confirm that Cassini detected nongyrotropic fluxes of outflowing CO$_2^+$ during both the R1 and R1.5 encounters.  Accounting for this nongyrotropy, we show that these possess comparable alongtrack densities of $\sim$2$\times$10$^{-3}$ cm$^{-3}$. Negatively charged pickup ions, also detected during R1, are surprisingly shown as consistent with mass 26$\pm$3 u which we suggest are carbon-based compounds, such as  CN$^-$, C$_2$H$^-$, C$_2^-$, or HCO$^-$, sputtered from carbonaceous material on the moons surface.  These negative ions are calculated to possess alongtrack densities of $\sim$5$\times$10$^{-4}$ cm$^{-3}$ and are suggested to derive from exogenic compounds, a finding consistent with the existence of Rhea's dynamic CO$_2$ exosphere and surprisingly low O$_2$ sputtering yields. These pickup ions provide important context for understanding the exospheric and surface-ice composition of Rhea and of other icy moons which exhibit similar characteristics.
 
\end{abstract}



\begin{article}

\section{Introduction} 

	Rhea is Saturn's largest icy moon with a radius of $\sim$764 km, and orbits within the sub-Alfv\'enic environment of Saturn's middle magnetosphere. As such, Rhea presents an archetype of the dominant satellite class at the outer planets, whose physical properties can be used to understand the formation and evolution of the giant planetary systems and, especially, their many moons.

	A sputter-induced exosphere was first discovered to exist at Rhea by the Cassini spacecraft \citep{Teolis10}, a phenomenon also present at Dione, Europa, Callisto, and Ganymede \citep{Tokar12,Hall98,Carlson00}. Rhea's and Dione's exospheres were also surprisingly found to host large quantities of CO$_2$ \citep{Teolis16}, a characteristic shared with Callisto. Rhea's surface is predominantly water ice whilst also containing lesser quantities of darker non-ice constituents and trace compounds such as CO$_2$ \citep{Clark08}.  Beneath this, Rhea’s gravitational field indicates a body existing away from hydrostatic equilibrium which might be differentiated \citep{Tortora16}, and possibly hosts a subsurface water ocean \citep{Hussmann06}.

\begin{figure*}
\includegraphics[width=1.0\textwidth]{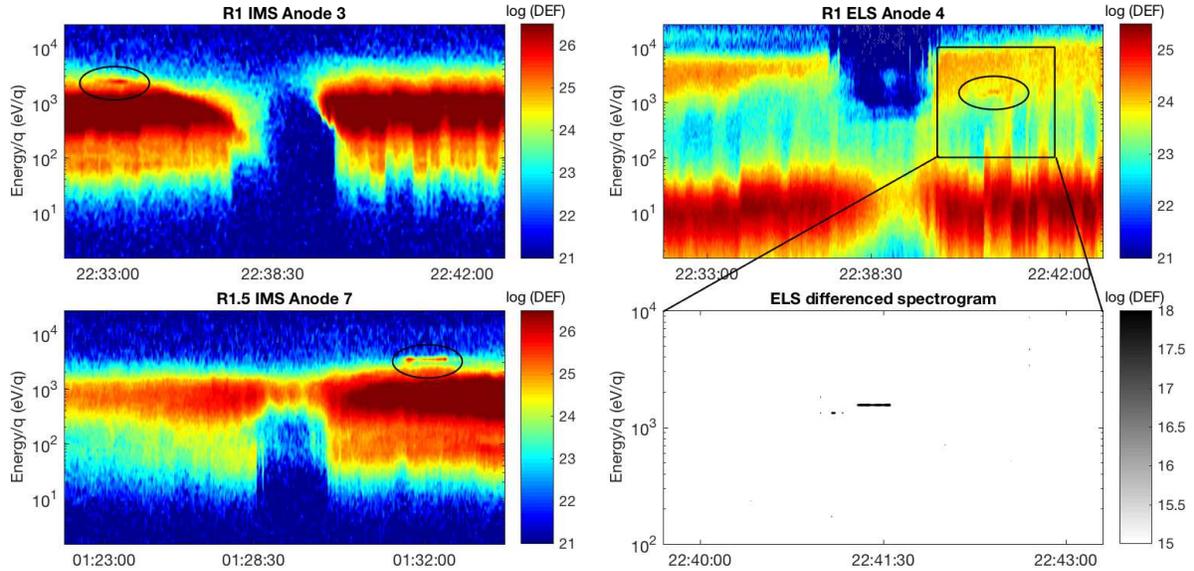}
\centering
\caption{The left-hand panels show CAPS IMS differential energy flux (DEF) spectrograms acquired during the R1 and R1.5 encounters with Rhea and the right-hand panels CAPS ELS DEF spectrogram acquired during the R1 encounter. The IMS pickup ion detections are encircled at 22:33:00 during R1 at $\sim$2.5 keV and 01:33:00 during R1.5 at $\sim$3.5 keV. The negative pickup ion detections are evident at 22:41:30 at $\sim$1.5 keV. The lower right-hand panel shows a differenced plot of the ELS data which shows a negative pickup ion signature similar in appearance to the positive pickup ion signatures.
\label{CAPS}}
\end{figure*}

	Rhea is an unmagnetised body and acts to absorb incident magnetodisk plasma \citep{Khurana08,Roussos08}. Ionised material can be directly picked up by the motional electric field and form ``pickup ion'' current systems which, with the resulting j$\times$B force and density gradients associated with the plasma wake \citep{Simon12,Khurana17}, slows down the incident magnetoplasma causing field-line draping and Alfv\'en wings. Pickup ions, as well as providing information on bulk and trace atmospheric constituents, impact the moon's plasma interaction and mass load Saturn's middle magnetosphere. 
	
	In a plasma flow, pickup ions will be accelerated to a maximum velocity of twice that of the bulk plasma and, in the plasma frame, will possess energies of
\begin{equation}
\label{PUI}
E_{i}=\frac{1}{2}m_{i}v_{b}^2sin^2\theta_{},
\end{equation}		
where $m_i$ is the pickup ion mass, $v_b$ is the bulk plasma velocity in the initial rest frame, and $\theta$ is the angle between the bulk plasma velocity and the magnetic field \citep{Coates89}. 

 At Rhea's orbit of $\sim$8.9 R$_S$, the Saturnian magnetic field is nominally dipolar and newly born ions will be accelerated perpendicularly to the magnetic field to execute rings in velocity space. If the size of the ion gyroradii significantly exceeds that of the pickup ion source region, the resultant distributions won't fill the entire ring and can be considered nongyrotropic. 

Pickup ion distributions are inherently unstable and provide a source of free energy for plasma wave generation \citep{Wu72}. These waves act scatter the distributions in pitch angle and energy, and heat ambient gyroresonant populations.

 Alfv\'en-cyclotron waves, generated by pickup ions, have been observed throughout Saturn's extended neutral cloud out to $\sim$8 Rs where the increased plasma beta (ratio of magnetic to thermal pressure) results in the Mirror Mode dominating \citep{Russell06,Meeks16}. The magnetic signatures of mass loading have, however, not been reported in the vicinity of Rhea, despite increased O$_2^+$ abundances being observed at these radial distances \citep{Martens08}.

In this \textit{Letter}, we examine Cassini Plasma Spectrometer (CAPS) observations of pickup ions outflowing from Rhea with emphasis on further constraining the composition and origin of the negatively charged pickup ions detected by the CAPS Electron Spectrometer (ELS).

\section{Velocity Space Analysis} 

The CAPS Ion Mass Spectrometer (IMS) and CAPS Electron Spectrometer (ELS)  \citep{Young04} were designed to measure low energy ions and electrons in the ranges of 1 eV to 50.3 keV and 0.6 eV  to 28.8 keV, respectively. CAPS is located on an actuator which was held fixed during the Rhea flybys.  Figure \ref{CAPS} shows the CAPS observations during the targeted R1 enounter on 26 November 2005 and the non-targeted R1.5 encounter on 30 August. Closest approach occurred at 765 km and 5736 km, respectively, and both flybys occurred behind the moon, thus providing the opportunity to observe outflowing material. 

During both encounters, a marked drop-out in ion and electron fluxes occurs as Cassini traversed the moon's plasma wake. During R1, distinct plasma populations are visible in the IMS spectrogram around 22:33 UT at $\sim$2.5 keV, and  during R1.5 a similar population is observed at 01:32 UT at $\sim$3.5 keV. In the ELS spectrogram, a distinct plasma population is visible at 22:41 UT at $\sim$1.6 keV, and a differenced plot, obtained by averaging the counts on anodes oriented away from 90$^{\circ}$ pitch angle (anodes 2, 3, 6, 7), and subtracting these from anode 3, reveals this signature as analogous to the IMS signatures. These respective plasma populations have been identified as positively and negatively charged pickup ions deriving from Rhea and provided evidence for the moon's tenuous exosphere \citep{Teolis10}.

\begin{figure*}
\includegraphics[width=1.0\textwidth]{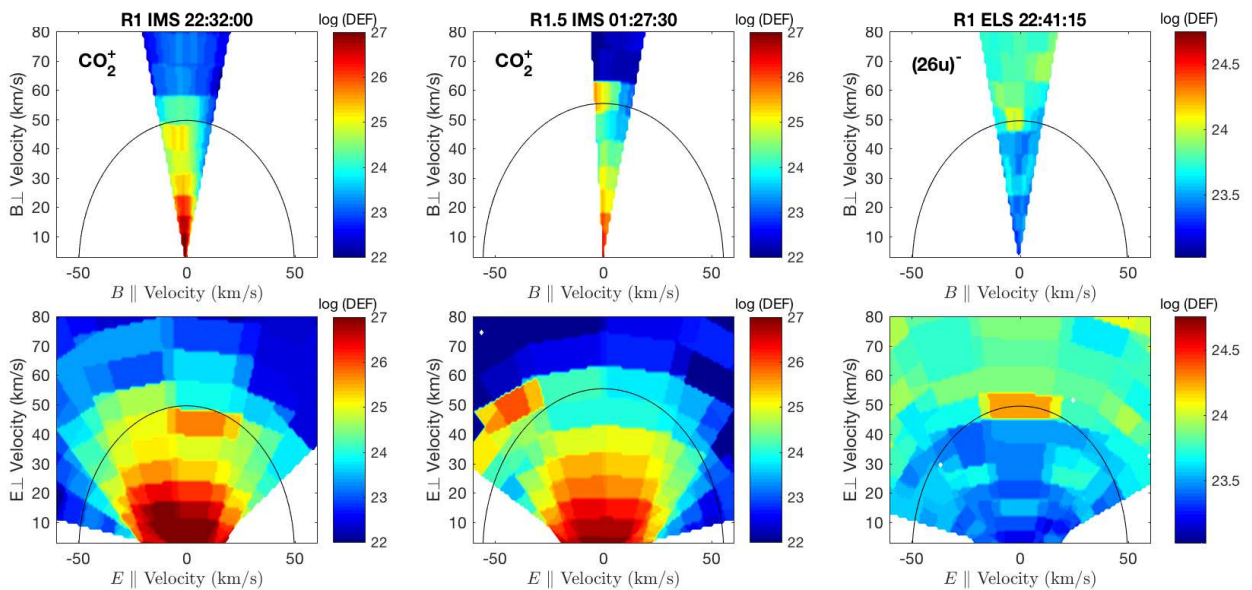}
\centering
\caption{The CAPS data are converted into velocity space assuming CO$_2^+$ (left and centre panels) and mass 26 u (right-hand panels). These are projected into planes parallel and perpendicular to the magnetic field and $-v \times B$ electric field for a single energy sweep. A linear interpolation is used for overlapping FOVs which dilutes the detections when projected relative to the magnetic field. The geometrical look directions are used which, combined with each anodes 20$^{\circ}$ width, smears the detections in the parallel and anti-parallel directions. The pickup velocity ring-contour are overlaid. 
\label{vspace}}
\end{figure*}

The IMS and ELS utilise electrostatic analysers to energy select charged particles, and the characteristic velocity imparted to newly created ions by the pickup process allows CAPS to discriminate between pickup ions of different masses.  Figure \ref{vspace} shows an IMS and ELS energy sweep corresponding to when these pickup ions were detected. The data are transformed into velocity space using the mass of anticipated pickup ions and projected onto planes which are parallel and perpendicular to the magnetic field and $-v\times B$ electric field, as measured by Cassini \citep[e.g.][]{Wilson10}. The spacecraft and plasma velocity are subtracted, leaving the measurements in the pickup ion rest frame and a contour representing the anticipated pickup ion ring distribution is overlaid, as predicted by Equation \ref{PUI}.

	During R1 and R1.5, the pickup ions observed by the IMS appear consistent with masses 40$\pm$4 u and 46$\pm$4 u, respectively. This uncertainty derives from the width of the IMS energy bins and the plasma velocity which varies between $\sim$55 and $\sim$60 km/s during R1, and $\sim$55 and $\sim$65 km/s during R1.5, see \citet{Wilson10}.  The pickup ions arrive with near-zero velocity parallel to the B-field as anticipated for pickup within a dipolar pickup geometry and are therefore attributed to CO$_2^+$, a conclusion previously reached by \citet{Teolis16}.
	
\begin{figure*}
\includegraphics[width=1.0\textwidth]{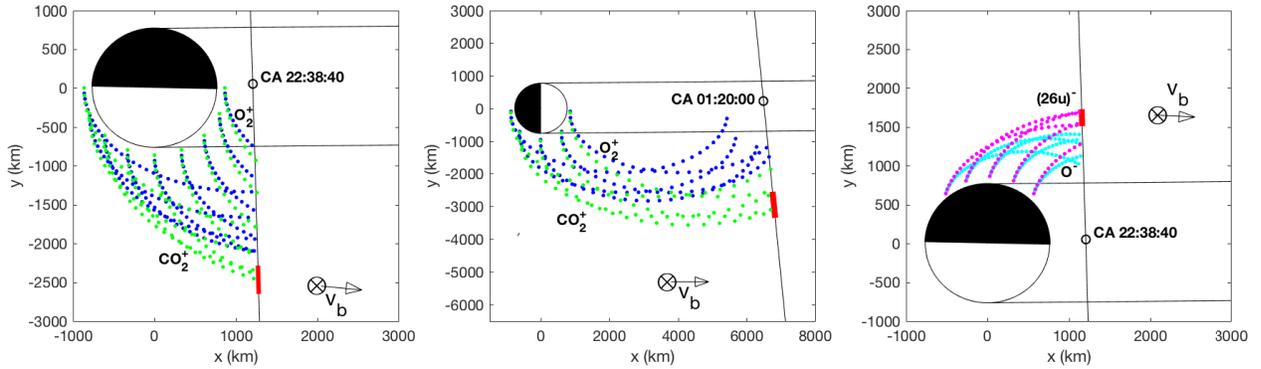}
\centering
\caption{Nominal trajectories of outflowing O$_2^+$ (blue) and CO$_2^+$ (green) during R1 in the left-hand panel and during R1.5 in the centre panels. Outflowing O$^-$ (cyan) and negative ions of 24u (magenta) are shown during R1 in the right-hand panel. The trajectories are calculated based upon Cassini plasma and field measurements at the time of detection and displayed in a Rhea centred coordinate system. The nominal corotational wake and approximate sunlit regions of Rhea are marked and times along Cassini's trajectory where the pickup ions were detected are marked red. The pickup ion trajectories originate within 100 km of Rhea's surface where increased neutral abundances are anticipated.
\label{gyrotrace}}
\end{figure*}

	The  pickup ions possess varying velocities parallel to the electric field indicating they are highly nongyrotropic and exist within slightly different locations in phase space. During both encounters the pickup ions appear shifted compared to the predicted velocity contours, the most likely explanation being that the plasma conditions were different closer to the moon, where and when the ions were produced, compared to at the time and location of their detection. In Figure \ref{gyrotrace}, the nominal trajectories of outflowing positively and negatively charged pickup ions are shown, calculated using Cassini field and plasma measurements. The CO$_2^+$ trajectories can be seen to correspond to where the respective detections were made during both R1 and R1.5. 

	The nongyrotropic nature of these distributions also becomes evident when examining these trajectories as they demonstrate how the pickup ions are only able to occupy a finite amount of the 2$\pi$ velocity-ring space at any given instance.  This also explains why the pickup ions are only observed over a finite time period as the pickup ion phase angle changes closer to the moon to where the CAPS finite FOV did not cover. Spatial variations in the ion production rate, due to the spatial distribution of the exospheric neutral density, could also contribute to this effect.

\begin{figure*}
\includegraphics[width=1.0\textwidth]{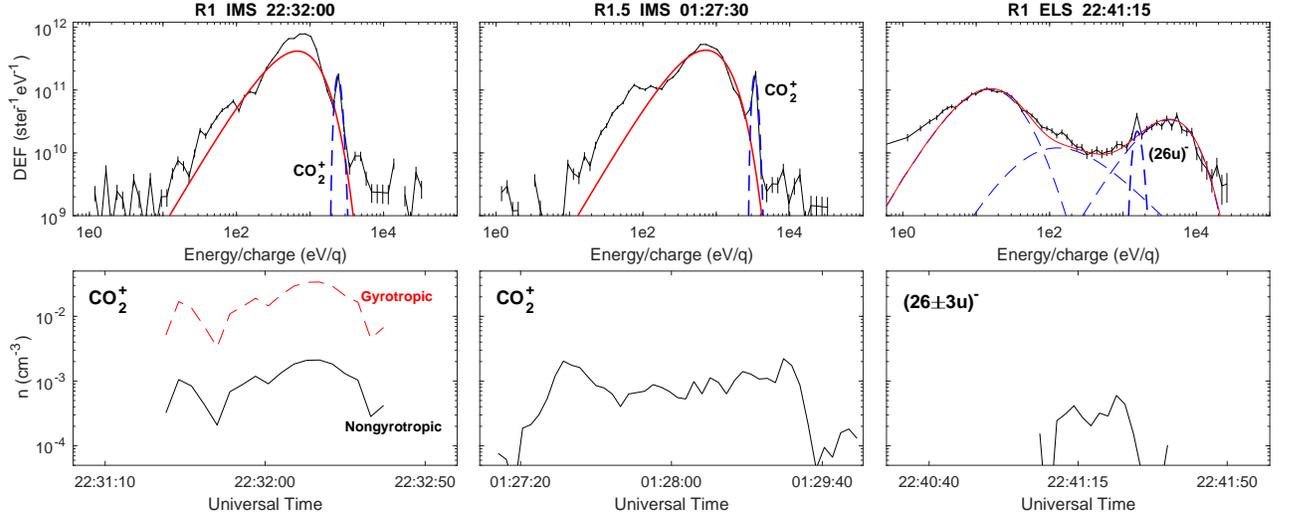}
\centering
\caption{The top panels show an IMS and ELS spectrum from each of the pickup ion detections identified in Figure \ref{CAPS}. The IMS background plasma is fitted to using Maxwellian velocity distributions for water-group ions and the ELS spectra is fitted using a low, medium and high energy kappa distributions. The lower panels show the calculated alongtrack densities. 
\label{CAPS2}}
\end{figure*}

	The identification of CO$_2^+$ on both the Rhea encounters raises interesting questions regarding the lack of O$_2^+$ pickup ions, as observed at Dione by Cassini \citep{Tokar12}. The IMS spectra shown in Figure \ref{CAPS2} do, however, feature a shoulder on the main corotational plasma distribution near $\sim$2 keV which, when closer to Rhea (not shown), appears similar to the O$_2^+$ pickup ion detections reported by \citet{Tokar12}. It is, however, difficult to differentiate this from the corotational plasma at Rhea's orbit, which exhibits a greater spread in energy compared to at Dione.	Rhea's CO$_2$ and O$_2$ exosphere has been measured insitu by Cassini's Ion and Neutral Mass Spectrometer \citep{Teolis16}, and the O$_2$ production rates have notably been determined to be significantly ($\sim$300 times) lower than that predicted from the sputtering of pure water-ice, and are consistent with the presence of significant surface impurities.
	
The consistency of these identifications with the analysis and exospheric modelling results reported by \citet{Teolis16}, validates the use of the pickup ions velocity as a means of identifying composition, a method which will now be applied to analyse the negatively charged pickup ions.

	The CAPS-ELS is capable of detecting negatively charged ions \citep{Coates07}, and \citet{Teolis10} reported that the negatively charged pickup ions detected during R1 were likely comprised of O$^-$. While initially reported as produced from electron attachment to atmospheric species, the inefficiency of this process was highlighted in a subsequent study \citep{Teolis16,Itikawa08}, and it was consequently suggested that these were likely produced by surface mediated process such as sputtering. In Figure \ref{CAPS} and \ref{vspace}, these detections appear above the anticipated O$^-$ energy by $\sim$15 km/s which corresponds to an energy discrepancy of $\sim$500 eV. These detections therefore appear consistent with a heavier species of mass 26$\pm$3 u. It is, however, possible for pickup ions to be accelerated to increased energies by a number of processes, which are now examined:

			\begin{itemize}
	
	\item Intense plasma waves have been observed at Rhea \citep{Santolik11}, and right-hand polarised Alfv\'en-cyclotron waves, which would gyroresonantly interact with O$^-$, could be produced by negatively charged pickup ions \citep{Desai17b}. These however grow from the free energy from the pickup ions and this effect could not be significant over such a short time period. 
	
	\item Specular reflection from the lunar surface has been observed to accelerate solar wind ions to three times that of the bulk plasma velocity \citep{Saito08}. This is however judged unlikely at Rhea due to high water-group photo-detachment rates in Saturn's magnetosphere \citep{Coates09} precluding negatively charged O$^-$ existing in abundance as an ambient magnetospheric population.  
	\item Sputtering can result in energy being transferred to the sputter products.  O$^-$ sputtering experiments have, however, shown this to be too inefficient to account for the velocity discrepancy discovered herein \citep{Tang96}. 
	
	\item Previous theoretical studies have predicted large negative surface potentials at Rhea up to several hundred volts [Roussos et al., 2010; Nordheim et al., 2014] and observations during the Rhea R2 flyby appear to support this \citep{Santolik11}. However, for surface potentials to explain the observed energy discrepancy, a negative surface potential of 500V would have to occur uniformly over a large region of Rhea's surface. Given that the theoretical studies have predicted surface potentials which vary strongly depending on surface location, this is not considered likely.
	
	\item The bulk plasma velocity is predicted to vary in the vicinity of Rhea and in particular on the Saturn-facing hemisphere \citep{Roussos08}. The CAPS plasma velocity measurements during R1 were, however, obtained in this region and  do not show this effect to be significant \citep{Wilson10}.  
	
	\end{itemize}  
  
	The apparent inconsistency with O$^-$ pickup ions raises two possibilities.  Firstly, the signature could be produced by an electron beam oriented perpendicularly to the magnetic field. 
	The longevity of this signatures above the background populations, its spatial occurrence, and the similarity to the unambiguous positive pickup ion signatures, are however highly indicative of negatively charged pickup ions of mass 26$\pm$3 u, of a type not previously considered.
	
\section{Origin of the Negative Ions} 

  Heavier negative pickup ions could result from carbon-based compounds with positive electron affinities (EA), such as CN$^-$, C$_2$H$^-$, C$_2^-$, or HCO$^-$, being produced via sputtering of the moon's surface. Spectroscopic observations of Rhea at $\leq$5.2 $\mu$m wavelengths have revealed unusually dark material which is consistent with the presence of either tholin (C-, H-, N-, O- bearing) and/or iron (Fe- bearing) compounds \citep{Ciarniello11,Stephan12,Scipioni14}. This material is also present at Dione, Phoebe, Iapetus, Hyperion, Epimetheus and throughout Saturn's F-ring, thus implying a common process occurring throughout these icy satellites \citep{Clark08}.  Dark tholin-like material is also apparent in spectroscopic observations of the Galilean icy moons, which is thought to be comprised of hydrocarbon or cyanide compounds \citep{McCord98}.  An abundance of electrons are also anticipated near Rhea's negatively charged surface which could readily attach onto electrophillic molecules. 
  
	Visual Infrared Mapping Spectrometer (VIMS) observations of Rhea have shown that an $\sim$1\% tholin-type admixture could explain unidentified features in the near-infrared \citep{Ciarniello11}.  It therefore initially appears surprising that such a trace constituent would be observed outflowing in significant quantities.  The pickup ion trajectories, however, shown in Figure \ref{gyrotrace}, demonstrate how pickup ions originating from different regions in Rhea's exosphere become concentrated in phase space when observed, a phenomenon which appears enhanced for heavier species due to their larger gyroradii. This could explain why a trace heavier species might preferentially be detected.  
	
	VIMS observations of Rhea's surface are also derived from the top few microns and exogenous material might not have penetrated this far.  A thin carbonaceous surface coating on the moon, possibly only a few monolayers thick, could therefore consist of significantly larger fraction of this unidentified dark material than is apparent from remote observations.  This could have been deposited from magnetospheric plasma and dust populations, or been delivered by micrometeorite, cometary and interplanetary particles raining into the Saturn system  \citep{Clark08,Stephan12}.  At Rhea, the spatial distributions of the unidentified material indeed suggest an external origin, with higher concentrations on the leading and trailing hemispheres pointing to magnetospheric dust and plasma deposition, respectively \citep{Clark08,Scipioni14}.  
	
	Dark spots are also observed near Rhea's equator which are associated with surface disruptions \citep{Schenk11}. The negative pickup ions map back to near Rhea's equatorial regions and the possibility that endogenic carbon-rich material could have been released onto the surface via impact events such as the cause of the Inktomi impact crater, or the discolored spots, or from further large-scale geologocial resurfacing \citep[e.g.][]{Stephan10}, cannot be discounted. 
  
  It is possible for tholin-type compounds to be incorporated into the ice from when the moon formed. Compounds such as C$_2$H$_2$, CH$_3$OH and HCN appear ubiquitously within cometary ices at $\sim$1\% of H$_2$O abundances, thus signifying their presence in the protosolar accretion disk from which the giant planetary systems formed \citep{Mumma11}.  \citet{Teolis16} go on to predict the quantity of carbon atoms in Rhea's surface ice to be as high as 13\% that of H$_2$O and processes such as photolysis, radiolysis, and heating could act to process the chemical state of such compounds, whether endogenic or exogenic to Rhea.

  The CN$^-$ (EA = 3.8 eV), C$_2$H$^-$ (EA = 3.0 eV) and C$_2^-$ (EA = 3.3 eV) anions can be produced from electrons impacting compounds such as hydrogen cyanide, acetylene and diacetylene \citep{Inoue66,May08}.  Graphite-compounds could be created through radiation bombardment and surface chemistry \citep{Lifshitz90,McCord98}, which could produce  C$_2^-$.  Sputtering experiments indeed predict the efficient production of the C$_2$H$_x^-$ anions from hydrocarbon compounds \citep{Johnson92}, and these are suggested as candidate sputter products at Europa \citep{Johnson98}. Carbon chain anions have also possibly been observed at Comet Halley \citep{Cordiner14} and exist elsewhere in Saturn's magnetosphere, amongst carbon-rich compounds in Titan's ionosphere \citep{Desai17a}. The HCO$^-$ (EA = 0.31 eV) anion could possibly be formed by deprotonation or dissociative electron attachment of H$_2$CO or CH$_3$OH.
  
  Further study of the sputtering of carbon-rich ejecta from ices representative of the icy moons of Saturn and Jupiter, could surely provide further insight into which sputtering rates are significant with regards to these anions.
 
	Although the aforementioned negative pickup ions appear inconsistent with O$^-$, it should be noted that in the differenced ELS spectrogram displayed in Figure \ref{CAPS}, further signatures also consistent with negative pickup ions are present earlier in time at a lower energy. Although too brief to conclusively identify these as negative ions, this might represent the same pickup ion population dispersed in energy via interactions with a surface-generated electric field as observed at the Earth's moon \citep{Poppe12}, or indeed could correspond to O$^-$ pickup ions given their location in phase space, see Figure \ref{gyrotrace}. 

 \section{Densities \& Escape Rates}  \label{densities}

	Figure \ref{CAPS2}	 shows IMS energy spectra during R1 and R1.5 and the ELS spectrum during R1, corresponding to the pickup ion detections. The ion background is fitted using a Maxwellian water-group velocity distribution which appears to better match the data during R1.5 when Cassini was farther from Rhea's plasma interaction. In either instance the pickup ion detections are clearly identifiable and these fits are sufficient to isolate the pickup ion populations.  A lower energy hydrogen population is also present, as well as further high energy populations evident at $>$4 keV. These are not considered for this analysis. 
	 
	 The ELS spectrum shows two distinct electron populations at low and high energies which are represented using a double Kappa distribution \citep{Schippers09}. A significant amount of intermediary electrons are also present which are also approximated by a broad Kappa distribution. The proximity to Rhea's plasma interaction, as well as multiple possible photoelectron populations \citep{Taylor17}, result in significant variabilities in the electron spectrum during R1.  The negatively charged pickup ion population consistently remains above the background throughout this variability, see Figure \ref{CAPS}. 
	 	  
	 The nongyrotropic pickup ion densities are calculated using an expression derived for partially-filled ring-velocity distributions. In the plasma frame, the pickup ions can be expressed relative to the magnetic field as
\begin{equation}
\label{ring}
f(v)=\frac{n}{\Delta\phi_{}v_{b \perp}} \delta (v_{perp}-v_{b\perp}) \delta (v_{\parallel}-v_{b\parallel}),
\end{equation}			 
where $\Delta\phi_{}=2\pi$ in the case of a gyrotropic ring \citep{Wu72}. 
	
	Pickup ion trajectories derived from the extrema of Rhea's exosphere, see Figure \ref{gyrotrace}, are used to estimate that the pickup ions are able to fill $\sim \pi/4$ of velocity space, see Figure \ref{gyrotrace}.  This can be expressed in the spacecraft frame as 
\begin{equation}
\label{ringsc}
f(v)=\frac{n}{\Delta\phi_{}v_{b \perp}} \delta (v_b-v_c) \delta v_{b\parallel},
\end{equation}	
where $v_c$ is the velocity corresponding to the CAPS energy bin in which the pickup ions were detected. The spacecraft velocity is removed for an inertial reference frame.  The pickup ion density, $n$, can then be calculated from the count rate, $R_c$ by the expression,
\begin{equation}
\label{density}
n=\frac{R_c \Delta \phi}{v_b \varepsilon A \Delta \phi_c} ,
\end{equation}	
where the area of acceptance $A=0.33 cm^{-2}$, the CAPS phase angle coverage  $\Delta \phi_c= \pi/2$, and $\varepsilon$ is the Microchannel Plate (MCP) efficiency. An MCP efficiency of 0.46 is used for the CO$_2^+$ pickup ions and 0.50 for the negatively charged pickup ions \citep{Tokar12,Stephen00}. 

Figure \ref{CAPS2} shows the resulting IMS and ELS along-track densities.  The CO$_2^+$ densities peak at $\sim$2.5$\times$10$^{-3}$ cm$^{-3}$ whereas the negatively charged pickup ions peak at values nearly an order of magnitude lower at $\sim$5x10$^{-4}$. The CO$_2^+$ densities during R1 are also calculated assuming a gyrotropic distribution to demonstrate how such an assumption significantly overestimates abundances. 

Whilst Rhea's exosphere has shown to be dynamic and variable, the CO$_2^+$  densities can be integrated over the hemisphere of the moon where the motional electric field will result in ion escape, to calculate approximate global escape rates. Assuming uniform ionisation, this results in an estimated $\sim$4.6$\times$10$^{20}$ CO$_2^+$ s$^{-1}$ escaping the moon during R1 and $\sim$5.7$\times$10$^{20}$ CO$_2^+$ s$^{-1}$ during R1.5. This rate is compatible with the varying CO$_2$ production rates resulting from the model of \citet{Teolis16}. This is also $\sim$0.25 times that predicted at Dione which experiences more intense plasma bombardment due to being located deeper inside Saturn's magnetosphere \citep{Wilson17}.  A similar calculation can be performed for the negative ions which results in an outflow rate of $\sim$5.4$\times$10$^{19}$ s$^{-1}$.  

It should be noted that these rates are only an estimate as studies have shown highly varying dynamical processes at Rhea's magnetospheric interaction at small or intermediate scales, and such dynamics may ``destroy" or disperse the smooth paths of pick-up ions with small or moderate-sized gyroradii \citep{Roussos12}.  It is therefore not clear precisely which density should be used to represent globally averaged ion production rates.

The outflowing pickup ions will contribute to magnetospheric ion populations. Rhea's CO$_2$ exosphere may therefore provide a source of the 44 u ions, or the carbon and oxygen ions via dissociative reactions, identified at radial distances of $<$20 R$_S$  \citep{Christon15}.  If the breakup is sufficiently fast, this may provide some explanation for the elevated O$_2^+$ levels reported by \citet{Martens08} at Rhea's orbit. 

Further analysis of the generation of instabilities associated with nongyrotropic pickup ions, in a plasma beta regime representative of Rhea's plasma environment, is required to understand whether the magnetic signature of this mass loading might be visible.

\section{Summary \& Conclusions}

	This study has analysed the composition, density and outflow rates of positively and negatively charged pickup ion distributions at Saturn's icy moon Rhea and determined the following:

	\begin{itemize}

 \item CAPS-IMS observed nongyrotropic fluxes of CO$_2^+$ pickup ions during the R1 and R1.5 encounters with comparable alongtrack densities of $\lesssim$2$\times$10$^{-3}$cm$^{-3}$.

  \item The R1 CAPS-ELS detections, previously identified as deriving from the pickup of O$^-$, are shown as consistent with a heavier species of mass 26$\pm$3u. These are consequently identified as negatively charged carbon-based compounds produced from tholin-type material on Rhea's surface.
  
   \item The negatively charged pickup ions are suggested to consist of CN$^-$, C$_2$H$^-$, C$_2^-$ or HCO$^-$, resulting from the dark material observed at Rhea and throughout the icy moons of Saturn.  
  
    \item The negatively charged ions were observed with alongtrack densities of $\lesssim$5$\times$10$^{-4}$ cm$^{-3}$.

       \item Possible further negative ion signatures are also identified which could represent dispersion in energy as a result of surface charging or a further population of O$^-$ pickup ions. 

\end{itemize}

This study provides context for understanding the exospheric and surface compositions and plasma interaction of Rhea as well as other icy satellites in the outer solar system.  The trace constituents in Rhea's surface ice, and also at other Saturnian and Jovian icy moons, are largely unconstrained and it remains to be determined just how similar or different these ices are to each other or indeed to ices formed elsewhere in the solar system such as those within comets and further icy bodies.


\begin{acknowledgements}
RTD acknowledges STFC Studentship No.1429777. AJC and GHJ acknowledge support from the STFC consolidated grants to UCL-MSSL ST/K000977/1 and ST/N000722/1. The Cassini CAPS data used are available on the Planetary Database System (PDS) or upon reasonable request. \end{acknowledgements}

\end{article}

\end{document}